\documentclass{IEEEtran}
\usepackage{capekCommands}
\usepackage{enumitem}

\usepackage{hhline}
\usepackage{booktabs}
\usepackage{arydshln}

\usetikzlibrary{shapes.geometric, arrows}

\newacro{iFFT}[iFFT]{inverse fast Fourier transform}
\newacro{iFT}[iFT]{inverse Fourier transform}
\newacro{FDTD}[FDTD]{finite differences in time domain}
\newacro{CST}[CST]{CST studio suite~\cite{CST}}
\newacro{QCQP}[QCQP]{quadratically constrained quadratic program}
\newacro{AToM}[AToM]{Antenna Toolbox for MATLAB~\cite{atom}}
\newacro{UWB}[UWB]{ultra-wideband}

\newcommand{\expn}{\T{e}}

\begin{document}

\title{Joint Signal and Topology Optimization \\ for Maximum Instantaneous Field Intensity}

\author{Jakub Liska, Lukas Jelinek, Miloslav Capek, \IEEEmembership{Senior Member, IEEE}
\thanks{Manuscript received \today; revised \today.}
\thanks{This work was supported by the Czech Science Foundation under project~\mbox{No.~24-11678S}, and by the Grant Agency of the Czech Technical University in Prague under project \mbox{No.~SGS22/162/OHK3/3T/13}.}
\thanks{J.~Liska, L. Jelinek and M. Capek are with the Czech Technical University in Prague, Prague, Czech Republic (e-mails: \{jakub.liska; lukas.jelinek; miloslav.capek\}@fel.cvut.cz).}}

\maketitle

\begin{abstract}
This paper introduces a computational approach to identify performance constraints in the time-domain, offering a way to design systems in pulse operation. 
This work presents a comprehensive application of convex optimization to determine fundamental bounds on time-domain waveforms.
The approach is applied to arbitrarily polarized multiport antennas and arrays, demonstrating their capability in maximizing peak radiation intensity in a specified direction and time under energy constraints.
This methodology allows us to consider matching, which is crucial in such applications.
To highlight the generality of the approach, receiving systems are also studied on an example of maximizing the local field for antiferromagnetic memory switching.
Thanks to its efficacy, this work enables joint optimization of excitation and system parameters.
\end{abstract}
\begin{IEEEkeywords}
Co-design,
optimal control,
excitation,
antenna topology,
pulse radiation,
peak power,
magnetic memory.
\end{IEEEkeywords}

\section{Introduction}
\IEEEPARstart{S}{hort} \ac{EM} pulse radiation is of significant importance in many fields, including applications involving impulse, short-pulse, or non-sinusoidal radars~\cite{1982_McIntosh_TAP,1986_Kang_TAP,Hackett2002}, ground penetrating radars~\cite{Maloney1993_OptimConicalAntPulseRad} or through-wall imaging radars~\cite{Yang2005WallRadar}.
In addition to radar applications, an important discipline within the field is \ac{EM} immunity testing~\cite{Chen1999_LPDA}.
Moreover, it has played a crucial role in communication systems that seek to increase information transmission rates~\cite{1982_McIntosh_TAP}, \ac{EM} pulse applications, broadband arrays and high-energy directed beam systems~\cite{1983_Pozar_TAP,1984_Pozar_TAP,1985_Pozar_TAP, Onder1992TAP_OptControlFeedVoltageDipAntenna}.
Apart from radioengineering and far-field operation, \ac{EM} pulses are used across the whole spectrum of physics, including antiferromagnetic switching~\cite{Olejnik201AntiferromagneticMemory}, which is analyzed in this manuscript.

Traditionally designed antennas using continuous wave analysis and synthesis techniques are frequently used to radiate short-duration pulses.
Antennas with wide bandwidth are typically used, relying on the belief that they ensure sufficient performance.
A more advanced approach involves matching an antenna excitation signal with its dispersion characteristics.
Such an approach yields significantly better results~\cite{1982_McIntosh_TAP,Etten1977,1983_Pozar_TAP}.
There are no widely recognized standards for evaluating the performance of an antenna that radiates pulses, although the following are generally accepted: the pulse distortion in the radiation should be minor, the energy of the reflected signal at the input should be small, or the amplitude of the radiated pulse should be large~\cite{Maloney1993_OptimConicalAntPulseRad}.

This paper establishes a computational approach to determine optimality in the instantaneous field intensity, generalizes the methodology, and allows the tool to be used in unconventional scenarios which had not been previously possible.
The tool can assess the quality of an arbitrary pulse system and give its optimal excitation.
The approach is coupled with system design process to reach better performance.

An essential advantage of the proposed method lies in its ability to assess whether a given system configuration can achieve the desired pulsed performance. The method provides critical insights into whether optimization efforts can be confined to pulse shaping or whether a redesign of the system geometry is necessary. This dual capability offers valuable guidance for both waveform optimization and structural design, facilitating the decision-making process.

A few works addressing optimal excitation in the time domain have been done, notably those considering radiated pulse optimization in~\cite{1984_Pozar_TAP,Polevoi_MaximumExtractableEnergy}, where up-to-date methodologies are not used.
Initial attempts to use modern tools in the optimization of radiated pulses are presented in~\cite{Jelinek_UpperBoundOnInstantaneousPowerFluxAPS23, Pozar2003_WaveformOptUWBRadioSys,Liska2023}.
These efforts, as well as this manuscript, build upon earlier developments of fundamental bounds in the frequency domain~\cite{GustafssonCapekSchab_TradeOffBetweenAntennaEfficiencyAndQfactor, GustafssonCapek_MaximumGainEffAreaAndDirectivity}, which established a foundation for waveform optimization using convex optimization techniques~\cite{BoydVandenberghe_ConvexOptimization}.

This paper demonstrates that the proposed method is adaptable to a variety of problems involving one or more sources using different simulation software. It also acts as a basis for automated co-design of pulse excitation and system topology.
The method and examples in the manuscript are organized as follows.
Section~\ref{sec:history} describes the history of how to generate excitation for optimal pulses.
The established computation framework is presented in Section~\ref{sec:comp}, and tested on an example of optimal directionally radiated pulses in Section~\ref{sec:verification}. A comparison of the performance of different antenna designs is shown in Section~\ref{sec:single}.
Using the same methodology, Section~\ref{sec:memory} reveals the optimal illumination that generates a strong localized electric field for antiferromagnetic memory switching, demonstrating its application in the local focusing and receiving parts of a system.
The co-design, including pulse and topology optimization, is presented on an antenna array with multiple ports, demonstrating the broad applicability of the approach later in Section~\ref{sec:codesign}.
Capabilities, limitations, and possible extensions are discussed in Section~\ref{sec:conclusion}.

\section{Historical Background}\label{sec:history}
This section recalls work done on optimal pulsed radiation, an oft-ignored topic in recent times, and highlights the novelty provided within this paper.

The critical advancement to uncovering the optimality of antenna pulse radiation lies in using an excitation signal precisely tailored to the specific dispersion characteristics of the antenna~\cite{1982_McIntosh_TAP}. Achieving this would result in a flat frequency response producing the highest possible pulse amplitude and maximizing the strength of the electric field at a given moment and point in the far field.
For arbitrarily shaped antennas, with a particular emphasis on planar pulse radiators, the search for optimality is detailed in~\cite{1982_McIntosh_TAP} and provides insight into performance maximization related to size and bandwidth.
The methodology assumes an arbitrary planar current distribution confined to the design region.
Although this understanding of bounds provides valuable insight, it falls short of synthesizing practical and applicable designs.

In~\cite{1983_Pozar_TAP}, a distinct performance maximum of pulse radiation is initially introduced to examine the optimal input voltage.
The metric and computational approach remain consistent. This exploits the computation in the frequency domain, using \ac{iFFT}, and applies variational optimization to maximize the amplitude of the radiated electric field at a specified time and position in the far field with constrained energy\footnote{Constraining the root mean square current amplitude over the surface of the radiator was implemented in~\cite{1982_McIntosh_TAP,1983_Pozar_TAP}.} and the bandwidth of the applied signal.
Compared to~\cite{1982_McIntosh_TAP}, the current in~\cite{1983_Pozar_TAP} is physically realizable, bringing this solution closer to practical implementation as long as the feed voltage can be generated or sufficiently approximated.
Although the solution in~\cite{1983_Pozar_TAP} refers to a dipole antenna, it accommodates arbitrary feed point locations, lengths, or impedance loadings.
In these works, the current distribution is determined using \ac{EFIE} and the Galerkin method to employ piecewise sinusoidal expansion~\cite{2012_Stutzman_Antenna_Theory}.

The progression of the study in determining the optimal transient radiation of arbitrary antennas is documented in~\cite{1984_Pozar_TAP, Pozar2007} and extends the analysis to three-dimensional antennas compared to planar radiators in~\cite{1982_McIntosh_TAP}.
The objective metric remains the same as before.
Furthermore, the study addresses the maximum possible radiated energy density within a specified time interval while considering the effects of introducing constraints on the sidelobes.
Physical realizability is emphasized by limiting Q or super-directivity effects.
Fundamental limitations are juxtaposed with the performance of optimally fed antennas~\cite{1983_Pozar_TAP,Pozar2007} and dipole arrays~\cite{1985_Pozar_TAP}, all of the same size~\cite{1984_Pozar_TAP,Pozar2007}.

Supplementary to~\cite{1983_Pozar_TAP} and in alignment with~\cite{1984_Pozar_TAP}, optimal excitations for dipole arrays aimed at maximizing the amplitude of the transient radiated electric field at a specified time and position in the far field are detailed in~\cite{1985_Pozar_TAP}, and maximizing energy density in a given time interval at a fixed far-field position are detailed in~\cite{1986_Kang_TAP}.
As with~\cite{1983_Pozar_TAP}, the \ac{MoM} is employed to calculate the currents on all dipoles, taking into account the effects of mutual coupling.

In addition to finding optimal antenna excitation, there is a search for optimal antenna design with regard to pulse radiation which cannot be accomplished using conventional antenna synthesis techniques~\cite{1982_McIntosh_TAP}.
This question is explored in~\cite{Maloney1993_OptimConicalAntPulseRad} with a focus on the optimality of conical antennas and in~\cite{Shlager1994_OptBowtieAntPulseRad} with an emphasis on bowtie antennas.
The theoretical analysis used to optimize the conical antenna is grounded in the time domain.
Geometrical parameters and resistive loading are adjusted to maximize radiated energy for a given excitation.
Similarly, dipole shapes are optimized~\cite{Wang1997_OptimDipoleShapes} using the conjugate gradient method.

Given that all previous work has focused on radiation in the far-field zone, the study of the focalization and hot spot creation remains open and can be addressed by the general framework in this manuscript.
Local focusing and concentrating \ac{EM} energy in small regions for short intervals is initialy studied and limitly addressed in~\cite{Hackett2002}.
Moving to communication systems, optimizing the transient waveform should consider both the transmitting and receiving antennas. In contrast to the work mentioned above, such an application can be covered by the approach proposed within this manuscript.
In \ac{UWB} radio systems, the crucial metric is the maximum amplitude of the received antenna voltage, its sharpness, or energy, as addressed in~\cite{Pozar2003_WaveformOptUWBRadioSys}.

Unbeatable target metrics are achieved by finding the best-performing waveform considering the constraints on energy in different stages of the channel.
General optimization results are derived for arbitrary antennas, considering the effects of the generator and load impedances using variational methods.
The approach is demonstrated with dipoles using \ac{MoM}\footnote{Piecewise sinusoidal basis is used as well, as in~\cite{1983_Pozar_TAP,1985_Pozar_TAP}.} solutions providing the necessary transfer functions and input impedances from the \ac{EFIE} analysis with the study enriched by closed-form results for short-dipole antennas.
Ideally, an \ac{UWB} transmitter's radiated power spectral density should be as flat as possible.
However, in reality, \ac{UWB} systems are limited by the performance of the antennas which is far from ideal~\cite{Pozar2003_WaveformOptUWBRadioSys}. To achieve higher performance, antenna--pulse co-design is demanded.

\section{Waveform Optimization}\label{sec:comp}
This section develops the general computational approach to maximize the instantaneous field intensity.
Various goals and objective functions with multiple constraints can be considered within the approach presented in this article.
The essential user input for the computation is the relation between the excitation and the target metric, \ie{}, a solution to Maxwell's equations.
Arbitrary linear scenarios can be studied, including optimization of radiated field intensity, power delivery efficiency, or waveform shaping tailored to specific spatial or temporal requirements.
The methodology developed in this Section ensures that globally optimal time domain excitation is found for a given system.

The selected objective function~$A(\M{a})$ should be either linear or quadratic in the \ac{dof}, such as instantaneous power flux at given point, and is optimized under a series of constraints, $B_i(\M{a})$, that are also quadratic or linear, notable examples being incident energy, delivered energy at a specific time frame, and field values at particular times and positions. The \ac{dof} are represented by incident power waves~\cite{Pozar_MicrowaveEngineering} collected in vector~$\M{a}$. In contrast to previous work, this methodology allows us to consider matching, which, as shown in subsequent sections, is essential for high performance in the pulse regime.

Mathematically speaking, the problem reads,
 \begin{equation} \label{eq:optim}
    \begin{split}
        \max \limits_{\M{a} \left( t \right)} \ & A(\M{a}), \\
        \T{s.t.} \ & B_i(\M{a}) \leq 0, \forall i \\
        & \omega \in \left[ \omega_\T{min}, 
        \omega_\T{max} \right].
    \end{split}   
\end{equation}
The pulse optimization problem is addressed in the Fourier domain for $\M{a} \left(\omega\right)$ by employing established techniques of convex optimization. The solution is further constrained by the available frequency band determined by the specific technology used in the design. Specifically, the angular frequency $\omega = 2\pi f$ and the frequency $f$ are considered within the range from $\omega_\T{min}$ to $\omega_\T{max}$, based on practical values\footnote{For example, in antenna radiation applications, the frequency band is limited using step-recovery diodes~\cite{Wang2023}, where $f_\T{min} = 0\,\T{Hz}$ and $f_\T{max} = 3.8\,\T{GHz}$.} for pulse generators.

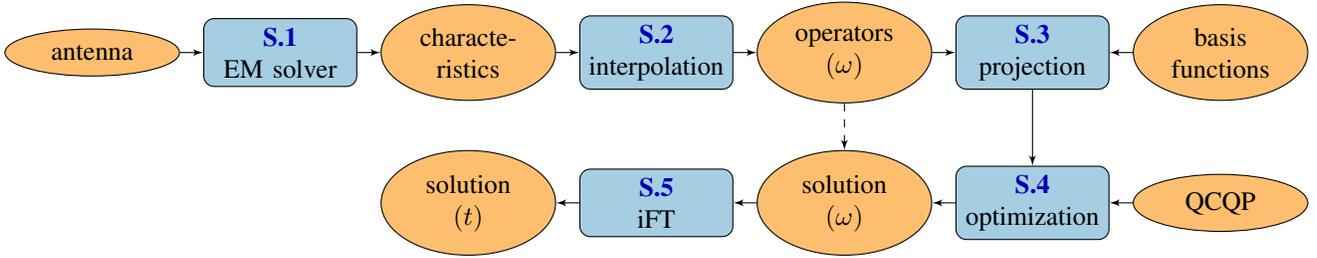
\begin{figure*}
    \centering
    \begin{tikzpicture}[node distance=2.5cm, auto]
\tikzstyle{block} = [rectangle, draw, fill=PairedA, text centered, rounded corners, minimum height=1em, minimum width=1.5cm]
\tikzstyle{line} = [draw, -latex']
\tikzstyle{cloud} = [draw, ellipse,fill=PairedG, text centered, text width=4em, minimum height=1em]

\node [cloud] (system) {system};

\node [block, right of=system] (solver) {\parbox{1.8cm}{\centering \ref{enum:solver} \\ \acs{EM} solver}};
\path [line] (system) -- (solver);

\node [cloud, right of=solver] (fline) {\parbox{4em}{\centering characte-ristics}};
\path [line] (solver) -- (fline);

\node [block, right of=fline] (interp) {\parbox{1.8cm}{\centering \ref{enum:interp} \\ interpolation}};
\path [line] (fline) -- (interp);

\node [cloud, right of=interp] (opf) {operators $(\omega)$};
\path [line] (interp) -- (opf);

\node [block, right of=opf] (proj) {\parbox{1.8cm}{\centering \ref{enum:proj} \\ projection}};
\path [line] (opf) -- (proj);

\node [cloud, right of=proj] (bf) {basis functions};
\path [line] (bf) -- (proj);

\node [block, below of=proj, yshift=0.5cm] (opt) {\parbox{1.8cm}{\centering \ref{enum:optim} \\ optimization}};
\path [line] (proj) -- (opt);

\node [cloud, right of = opt] (QCQP) {QCQP};
\path [line] (QCQP) -- (opt);

\node [cloud, left of=opt] (solf) {solution $(\omega)$};
\path [line, dashed] (opf) -- (solf);
\path [line] (opt) -- (solf);

\node [block, left of=solf] (ft) {\parbox{1.8cm}{\centering \ref{enum:iFT} \\ \acs{iFT}}};
\path [line] (solf) -- (ft);

\node [cloud, left of=ft] (solt) {solution $(t)$};
\path [line] (ft) -- (solt);

\node [block, left of=solt] (redesign) {\parbox{1.8cm}{\centering \ref{enum:redes} \\ redesign}};
\path [line, dashed] (solt) -- (redesign);
\path [line] (redesign) -| (system);

\end{tikzpicture}
    \caption{Flow chart of the computation steps~\ref{enum:solver}~--~\ref{enum:redes}, starting from the given antenna and port with the characteristic impedance of the feed transmission line and finishing with the optimal solution in the time domain, \ie, the optimal excitation and performance value which may show a need of redesigning the system.}
    \label{fig:flowChart}
\end{figure*}

 The problem can be tackled by taking and performing the following steps which are also depicted in the flow chart in Fig.~\ref{fig:flowChart}.
\begin{enumerate}[label=\textbf{S.\arabic*}]
    \item \label{enum:solver} 
    First, a system with defined excitation forms and input channels must be selected. The setup is then analyzed using an arbitrary \ac{EM} solver capable of computing the resulting field intensity at a specified location for a given frequency and a set of incident power waves. The solver provides a set of field values corresponding to different excitations.
    An arbitrary method, capable of evaluating target quantities for a given excitation, can be used, \ie, \ac{FDTD} in \ac{CST}. This part of the process is computationally the most time-consuming. The computational demands grow with wider frequency ranges $\omega \in \left[ \omega_\T{min}, \omega_\T{max} \right]$, but these, on the other hand, introduce more \ac{dof} in the waveform design providing greater flexibility in shaping the pulses.

    The provided field vector\footnote{It is not strictly required to focus on fields; instead, parameters such as voltage at a port, outgoing power waves along a transmission line, or other related metrics can be analyzed. The crucial aspect is the linear relationship between these metrics and incident waves, as described by the transfer matrix.} expressed in the frequency domain reads
    \begin{equation}\label{eq:Etarget}
        \V{E} (\V{r},\omega) = \M{H}(\V{r},\omega) \, \M{a}(\omega),
    \end{equation}
    where~$\M{H}$ is a transfer matrix obtained from the chosen solver.
    An example of getting the transfer matrix is the combination of the \ac{EFIE} and the \ac{MoM} detailed in Appendix~\ref{app:MoM}.
    Apart from the quantities needed for optimization, system properties can be examined to provide insight into the spectral selectivity of the analyzed system.
    
    \item \label{enum:interp} The field data are interpolated using rational fitting (sparse data) or linear interpolation (dense data) to provide a smooth function of frequency.
    In this step, the operator for energy and field is prepared, the operators are listed in Appendix~\ref{app:operators}.

    \item \label{enum:proj} The spectral functions representing the necessary operators for optimization, included in the objective function and constraint are projected onto a set of continuous-band-limited basis functions.
    The projection follows Galerkin method~\cite{Kantorovich1982,Harrington_FieldComputationByMoM}.
    The set of basis functions~$\xi_n(\omega)$ is described in Appendix~\ref{app:BF}.
    For all considered examples, number of basis functions\footnote{Later, the number of basis functions is the number of \ac{dof} in the optimization and determines the time needed for construction $\mathcal{O}(N^2)$ of matrices and solution $\mathcal{O}(N^3)$.} $N \leq 120$ was sufficient and utilizing additional basis functions does not change the result.
    The projection results in expansion coefficients collected in column vector~$\M{q}$, where   
    \begin{equation}\label{eq:expan}
    a_p(\omega) = \sum \limits_n q_{pn} \xi_n(\omega).
    \end{equation}
     The vector is constructed by blocks, each block representing one source.
 This allows to express functionals from Appendix~\ref{app:operators} as quadratic forms in vectors~$\M{q}$.
    Taking energy as an example, the quadratic form reads
    \begin{equation}\label{eq:Wmat}
        W(\V{r}, T) = \M{q}^\herm \M{W} (\V{r}, T) \M{q}
    \end{equation}
    with matrix elements $\M{W}_{pm,qn}$
where $pm,sn$ corresponds to the matrix rows and columns according to sources~$p,q$ and basis function~$m,n$, respectively.
    
    \item \label{enum:optim} The expansion coefficients are optimized to maximize the target performance metric according to~\eqref{eq:optim} which is, using projections described in Step~\ref{enum:proj}, transformed into a solution to \ac{QCQP}\footnote{In later examples equality constraint is used, because the optimum is always reached with the equality.}
    and reads
     \begin{equation} \label{eq:optimQCQP}
    \begin{split}
        \max \limits_{\M{q}} \ & \M{q}^\herm \M{A} \M{q} + \RE{\M{a}_i^\herm \M{q}} + \alpha, \\
        \T{s.t.} \ & \M{q}^\herm \M{B}_i \M{q} + \RE{\M{b}_i^\herm \M{q}} + \beta \leq 0, \forall i
    \end{split}
    \end{equation}
    where one of the matrices $\M{B}$ must be full rank and positive definite. This one is typically the total incident energy.
    
    The optimization task is resolved using standard methods of the \ac{QCQP}.
    The optimization problem is convex, so the solution is unique~\cite{BoydVandenberghe_ConvexOptimization}, which ensures that the maximum is truly achieved\footnote{The optimization problem is equivalent to that for determining performance limitations on antenna metrics in the frequency domain~\cite{Liska-CompFunBoAntennas-EuCAP22}.}.
    Within this paper, the optimization is more than a thousand times faster than the computation of the impulse response of the system (depending on discretization in spatial and frequency or domain).
    For more details about the convex optimization framework, see~\cite{Liska_etal_FundamentalBoundsEvaluation}.
    
    \item \label{enum:iFT} The solution to~\eqref{eq:optimQCQP} is found as optimal vector~$\M{q}$. The desired optimal excitation is then found as 
    \begin{equation}
    a_p(t) = \sum \limits_n q_{pn} \xi_n(t),
    \end{equation}
    where inverse Fourier transforms of basis functions~$\xi_n(\omega)$ are known.

    \item \label{enum:redes} If the performance of the current system with the optimal excitation is not sufficient, a redesign is needed, since solely by excitation, the performance cannot be enhanced further. For that, automated computer-powered system optimization can be employed. Automated co-design of the excitation and system topology then allows for even better performance.

    The automated co-design is computationally demanding and is realistic only when globally optimal excitations are found quickly. This is the case of a solution to \ac{QCQP} developed in previous steps.
\end{enumerate}

\section{Pulse Optimization Verification}
\label{sec:verification}
The steps mentioned above are applied to an example of a single dipole antenna  and instanenous far-field intensity~$\V{F}(\V{d},t)$ in direction~$\V{d}$ orthogonal to the dipole orientation.
Since this serves as verification, the redesign step~\ref{enum:redes} is omitted, and codesign loop is presented in Section~\ref{sec:codesign}.
The calculation is simultaneously performed in the \ac{AToM} and \ac{CST}.
The dipole with a length of $L=150\,\T{mm}$ and a width of $3\,\T{mm}$ is oriented along the $z$ axis.
The port is placed in the middle of the dipole and is connected to an ideal $50\,\T{\Omega}$ feed line.
Radiation in the spherical coordinate system is studied in the direction $\varphi = 0$ and $\theta = \pi/2$.
Far field consists of polarizations $\varphi$ and $\theta$.
Azimuthal coordinate $\varphi$ references to angle from positive direction of $x$ axis in $xy$ plane and elevation coordinate $\theta$ references to angle from positive direction of $z$ axis.

\begin{figure}
    \centering
    \includegraphics{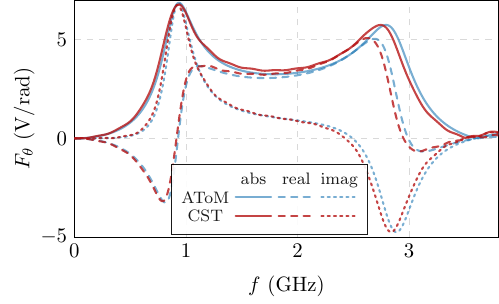}
    \caption{Far field along $\theta$ direction $F_\theta$ of a dipole antenna with a length of $15\,\T{mm}$ and width of $3\,\T{mm}$ in directions $\varphi = 0$ and $\theta = \pi/2$.
    The antenna is fed through a delta-gap source connected to the feed line, providing a voltage of $1\,\T{V}$ across the entire frequency spectrum.}
    \label{fig:ff}
\end{figure}

The antenna, including the port, is given with frequency range from $0\,\T{GHz}$ to $3.8\,\T{GHz}$~\cite{Wang2023}, and all steps visualized in Fig.~\ref{fig:flowChart} can be performed.
Choosing \ac{AToM} and \ac{CST} as the \ac{EM} solvers and computing the far-field along $\theta$ direction $F_\theta$ in elevation direction for unity excitation at all frequencies results in Fig.~\ref{fig:ff}.
Since the dipole is aligned along the $z$-axis, only the $\theta$ component of the electric field corresponds to radiation and far-field along direction $\phi$ is vanishingly small.
The curves are in good agreement.

In addition to the far field, an essential characteristic is total incident energy.
This step accomplishes \ref{enum:solver}.

Since the frequency data from \ac{CST} are dense due to the 1001 samples, only linear interpolation is used. Data resulting from \ac{AToM} are sparse and rational fitting is used for interpolation\footnote{Vector fitting~\cite{1999_Gustavsen_TPD} could be applied to enforce passivity and causality.} according to \ref{enum:interp}.

Metrics such as far field and total input power are projected to the band-limited functions defined in Appendix~\ref{app:BF} resulting in matrices as denoted in \ref{enum:proj},
\begin{equation}
    \V{F}(\V{d},t) = \M{F} \M{q}.
\end{equation}
The matrices allow for optimization~\ref{enum:optim} which gives the optimal solution.

\begin{figure}
    \centering
    \includegraphics{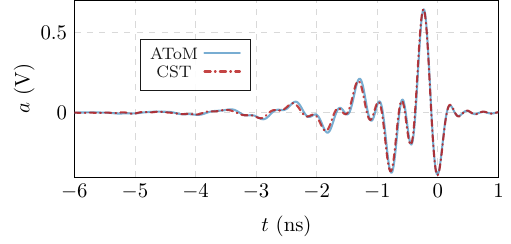}
    \caption{Optimal waveform of the incident wave gained from analysis in different solvers.
    In both cases of \ac{AToM} and \ac{CST}, optimization is performed in MATLAB.}
    \label{fig:at}
\end{figure}

Thanks to basis function orthonormality, the identity matrix of the incident energy is used as the full rank positive definite matrix in the \ac{QCQP},
\begin{equation}\label{eq:maxE}
    \begin{split}
        \max \limits_{\M{q}} \ & \M{q}^\herm \M{F}^\herm \M{F} \M{q}, \\
        \T{s.t.} \ & \M{q}^\herm \M{q} = W_0.
    \end{split}
    \end{equation}
The optimal excitation is a complex conjugate of the far field~\cite{1982_McIntosh_TAP} in Fig.~\ref{fig:ff}.
For presentation purposes, the total input energy was set to $W_0 = 10^{-10} \, \T{J}$.
Using~\ref{enum:iFT}, the optimal waveform in the time domain is plotted in Fig.\ref{fig:at}.
The agreement of different solvers is more than satisfactory.

\begin{figure}
    \centering
    \includegraphics{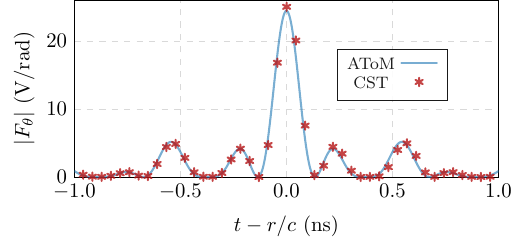}
    \caption{Far-field along $\theta$ direction $F_\theta$ optimal radiated waveform for $t_0 = 0 \, \T{s}$ at distance $r$ gained from \ac{iFT} of the far field for \ac{AToM} data and monitoring the far field during simulation in \ac{CST}, where the optimal incident wave is applied as excitation.}
    \label{fig:ft}
\end{figure}

The last step is the computation of the radiated pulse, shown as the magnitude of far-field along $\theta$ direction $F_\theta$ in Fig.~\ref{fig:ft}.
The similarity of the data computed by \ac{AToM} with \ac{iFT} and a synthetic measurement of the radiated pulse by optimal waveform excitation in \ac{CST} verifies and, even more, generalizes the method for determining optimal pulses studied for decades~\cite{1983_Pozar_TAP,1985_Pozar_TAP,Pozar2007,Jelinek_UpperBoundOnInstantaneousPowerFluxAPS23,Liska2023}.
In addition, this framework allows us to find an optimal excitation using an arbitrary solver and do it for complex scenarios, as demonstrated in the following section.

\section{Far-Field Radiation}\label{sec:single}
A representative example of a transmitting system is a single port antenna radiating in the target direction in the far field.
The methodology can be used to find the one with the best performance, \ie, with the highest allowed intensity of the instantaneous field.
As has been concluded in previous work on pulsed farfield radiation, the bowtie dipole antenna exhibits good performance among planar antennas studied for pulsed radiation in the direction orthogonal to the design plane~\cite{Jelinek_UpperBoundOnInstantaneousPowerFluxAPS23,Liska2023}.

In this section, the following antennas are considered: a self-grounded bowtie antenna, a conical dipole antenna, a conical spiral antenna, and a logarithmic periodic antenna.
The optimal waveform is determined to maximize the radiated peak power.
Incident power waves are used as degrees of freedom, which allows us to consider matching (reflected power waves).
This is essential for a realistic pulse response.
This example aims to demonstrate the utility of the approach and compare the best potentially achievable performance of the proposed antennas, including their optimal excitation waveform.

All samples can be circumscribed by the smallest sphere with a radius of $ a=36\,\T{mm}$.
The allowed frequency band of the input signal is chosen between~$0\,\T{GHz} - 15\,\T{GHz}$.
The antennas are simulated as if made of copper and connected to the $50 \, \Omega$ feeding line.
The observation point of the pulse radiation is always chosen in a direction native to each antenna.

\begin{figure}[t]
    \centering
    \includegraphics[width=0.25\textwidth]{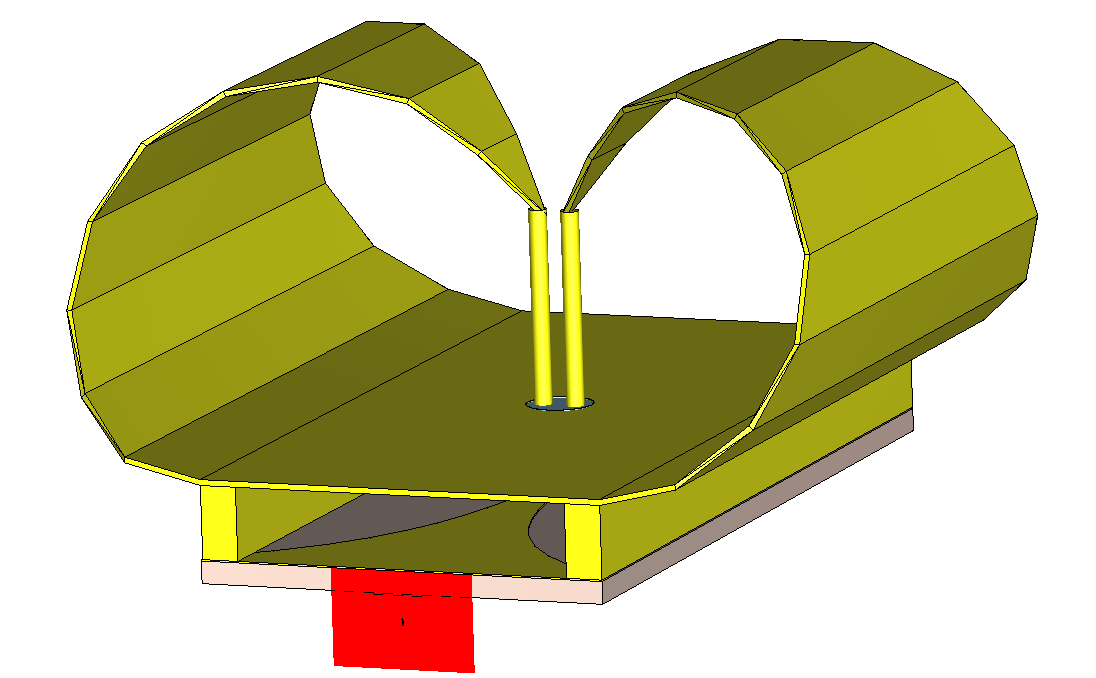}
    \caption{Self-grounded bowtie antenna with matching circuit as described in~\cite{Yang2012GroundedBowtie}.}
    \label{fig:groundedBowtie}
\end{figure}

The self-grounded bowtie is taken as proposed in~\cite{Yang2012GroundedBowtie}, the model is created and simulated in \ac{CST} and can be found in Fig.~\ref{fig:groundedBowtie}.
The authors~\cite{Yang2012GroundedBowtie} ascribe great performance in pulsed radiation to this format.

\begin{figure}[t]
    \centering
    \includegraphics[width=0.25\textwidth]{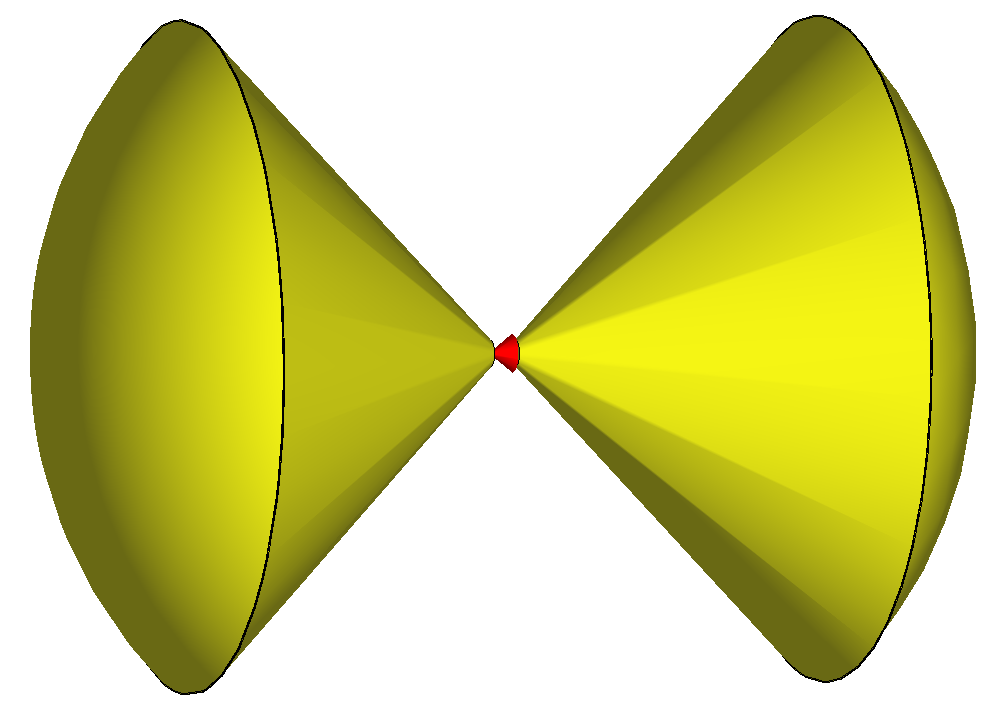}
    \caption{Volumetric bowtie dipole antenna.}
    \label{fig:bowtie3d}
\end{figure}

A planar version of the bowtie antenna is analyzed in~\cite{Jelinek_UpperBoundOnInstantaneousPowerFluxAPS23,Liska2023} in relation to pulsed radiation.
To take advantage of the volume of the circumscribed sphere, a volumetric bowtie dipole antenna is considered and scaled to fit the sphere.
See Fig.~\ref{fig:bowtie3d}, which is taken from \ac{CST}, where the simulation is performed.

\begin{figure}[t]
    \centering
    \includegraphics[width=0.4\textwidth]{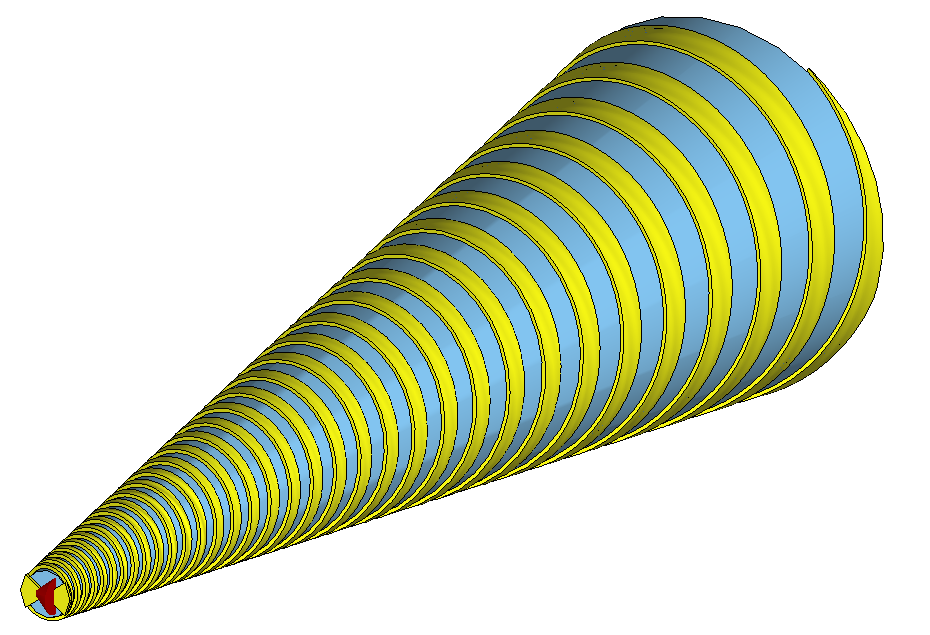}
    \caption{Conical spiral antenna for ground penetrating radar~\cite{Hertel2002ConicalSpiral}.}
    \label{fig:conicalSpiral}
\end{figure}

In~\cite{Hertel2002ConicalSpiral}, the conical spiral antenna is proposed for ground-penetrating radars, an \ac{UWB} application.
The conical spiral radiates a circularly polarized field compared to the other designs which operate in a linearly polarized regime.
The antenna proposed in~\cite{Hertel2002ConicalSpiral} is scaled to fit the circumscribing sphere mentioned above.
The simulation is again performed in \ac{CST} and the antenna is shown in Fig.~\ref{fig:conicalSpiral}.

\begin{figure}[t]
    \centering
    \includegraphics[width=0.45\textwidth]{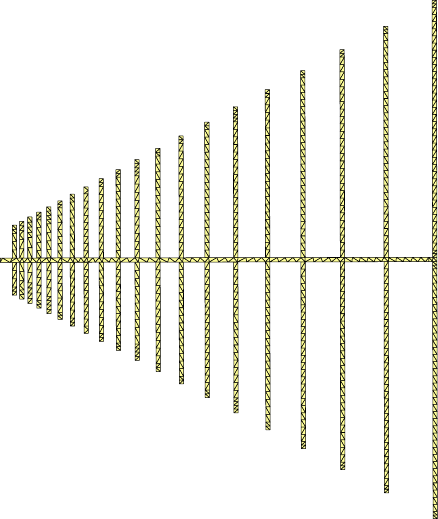}
    \caption{Logarithmic-periodic dipole wire antenna designed according to~\cite{Balanis1989}.
    The antenna is discretized in the mesh of $1528$~triangles.}
    \label{fig:logper}
\end{figure}

The periodic logarithmic antenna, recognized as an \ac{UWB} antenna~\cite{Balanis1989}, is extensively utilized in electromagnetic emission and immunity testing~\cite{Chen1999_LPDA}.
Therefore, a wire dipole array is a potential candidate for pulsed radiation and is compared with other antennas.
The array is designed to fit the circumscribing sphere and exhibits a decay of~$0.9$ in element lengths and an opening angle of $15$ degrees.
The strip width equals $0.017$ times the length of the longest element and the number of elements is~20.
The paremeters are manualy tuned to achieve better performance.
The antenna is simulated using \ac{AToM} and shown in Fig.~\ref{fig:logper}.

\begin{figure}
    \centering
    \includegraphics{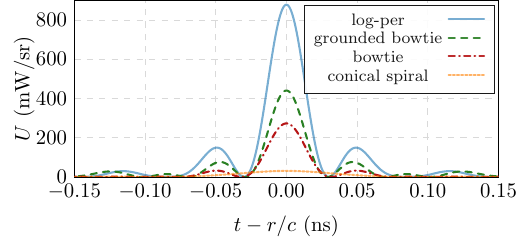}
    \caption{Optimal radiation intensity $U$ at chosen distance~$r$ from the compared antennas.}
    \label{fig:tSrad}
\end{figure}

The question remains as to which antenna performs better.
A comparison of radiation intensity~\cite{Balanis_Wiley_2005}

\begin{equation}\label{eq:Srad}
    U (t) = \dfrac{1}{Z_0} \left| \V{F} (t) \right|^2
\end{equation}
is shown in Fig.~\ref{fig:tSrad}.
The logarithmic periodic antenna exceeds the other designs in performance.
The performance of the self-grounded bowtie antenna is about two times lower and the other designs perform even worse.
This finding could prompt efforts to develop improved designs. This may be achieved through redesign, involving iterative tuning of the logarithmic-periodic dipole wire antenna. An advanced topology optimization way is shown in Section~\ref{sec:codesign} using an automated approach on a planar antenna.

\begin{figure}[t]
    \centering
    \includegraphics{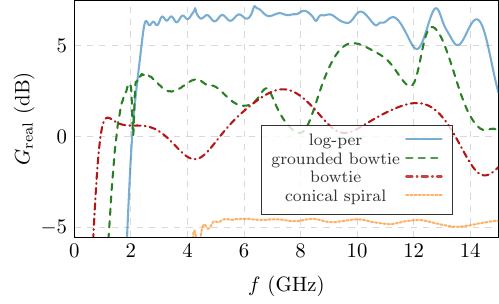}
    \caption{The realized gain $G_\T{real}$ of the antenna, which incorporates both the antenna gain and the effects of impedance matching, is a critical factor for achieving optimal performance.}
    \label{fig:Greal}
\end{figure}

Regarding the origins of what constitutes good performance, the realized gain in the given direction accounts for the pulse's sharpness and energy.
The emerging differences in realized gain $G_\T{real}$ of the antennas are depicted in Fig.~\ref{fig:Greal} which justifies the performance differences presented in Fig.~\ref{fig:tSrad}.
The highest and flattest realized gain gives the best performance which provides some direction for improved future designs. It is also noted that matching efficiency is a major part of the good realized gain of the logarithmic periodic antenna.

\begin{figure}[t]
    \centering
    \includegraphics{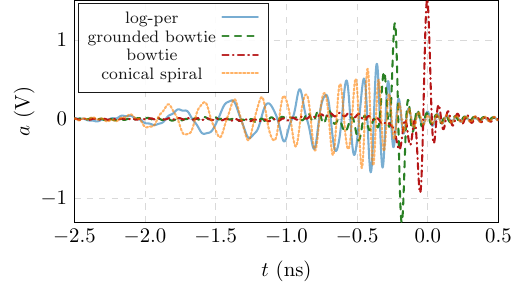}
    \caption{Optimal excitation signals achieving the highest performance in pulsed radiation for individual antennas.}
    \label{fig:aSrad}
\end{figure}

In contrast, phase properties must be considered once an optimal excitation signal is generated.
Abrupt changes in phase require more complex excitation than antennas with linear phase progress which prefer short excitation waveforms. 
The optimal excitation waveforms shown in Fig.~\ref{fig:aSrad} illustrate the ability of the approach to determine the optimal excitation for an arbitrary setup and how realistic it is to achieve the best performance of the given antennas.

The significant difference in excitation signals is visible for the bowtie and periodic antennas.
Sharp excitation is preferred by bowtie antennas as compared to periodic antennas which require a longer waveform for pulsed radiation.
The delay in the antennas is also noteworthy as the bowtie antenna immediately radiates energy, whereas the grounded bowtie excitation signal shows a delay in the matching circuit.
Meanwhile, the periodic antennas, namely the logarithmic and conical spiral ones, accumulate energy first.

The antennas considered in this section, such as the self-grounded bowtie, log-periodic, and dipole arrays, are selected as representative examples of broadband antenna designs. These types of antennas are often referenced in the literature for pulsed or time-domain applications because of their wide operational bandwidths and relatively flat frequency responses. In particular, the self-grounded bowtie antenna is highlighted in~\cite{Yang2012GroundedBowtie} for its advantageous broadband impedance characteristics and compact geometry, which are desirable for radiation in the time domain. However, the results presented in Fig.~\ref{fig:tSrad} reveal that while these characteristics make the bowtie antenna a practical choice, its instantaneous field intensity performance in the evaluated scenario falls short of that of the log-periodic antenna. This illustrates the utility of the proposed framework in identifying and quantifying such performance gaps and in guiding design improvements for pulsed applications.

The comparison of the antennas and the finding of the optimal excitation using the approach proposed in this paper show that antenna design and pulse generation capabilities are closely related.
The information that can be provided by this computational approach can provide essential insight into designing antennas that perform well in pulsed radiation. 

Notably, within the optimization framework~\eqref{eq:optim}, constraints on field intensity at spatial points can be included without compromising generality or convexity. Similarly, upcoming sections demonstrate the imposition of energy constraints at the target point within a time window. This method can also be extended to additional spatial points outside the main direction and time frames, still using the same framework.  However, the inclusion of these constraints may result in an increase in the duration of the optimization process.

\section{Field Focusing}\label{sec:memory}
Apart from far-field transmission, the focalized received power, including the performance of hotspot creation in a medium, is also covered by the approach.
This example illustrates field focusing, where an incident wave interacts with a large electrical structure, producing a localized and intense electric field pulse.

The methodology is demonstrated on a challenging example of antiferromagnetic memory switching using terahertz laser pulse~\cite{Olejnik201AntiferromagneticMemory}.
The goal is to reach a peak electric field intensity high enough for memory switching. Although the switching process is non-linear, here we only solve the field creation, which is a linear process. 

\begin{figure}[t]
    \centering
    \includegraphics[width=0.25\textwidth]{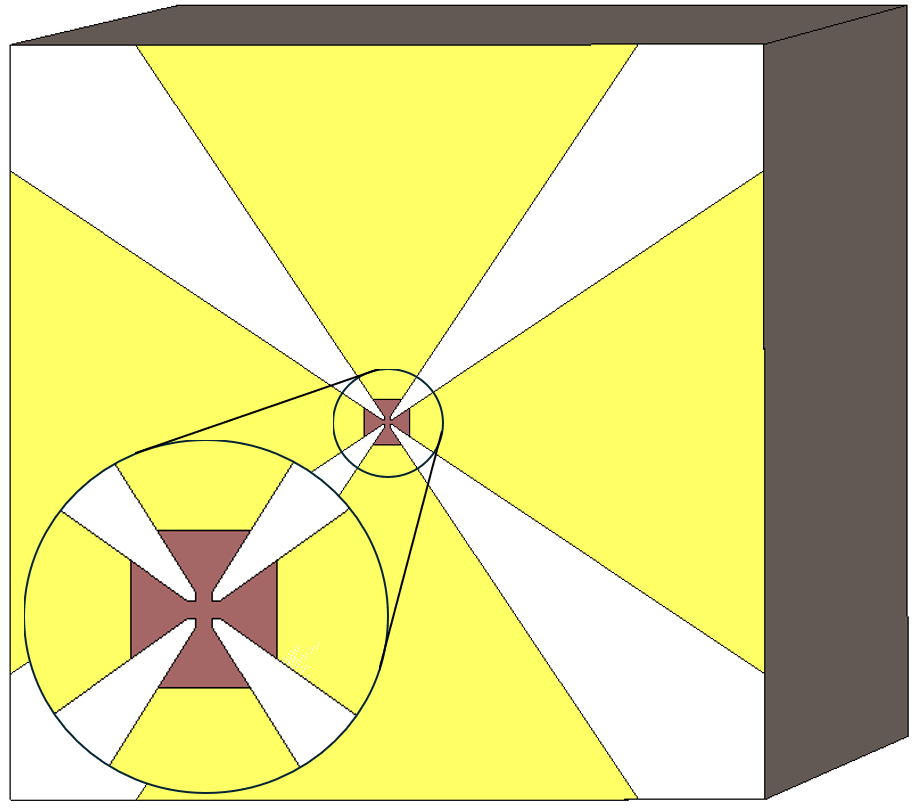}
    \caption{\ac{CST} model of the central part of the antiferromagnetic memory~\cite{Olejnik201AntiferromagneticMemory} with zoomed target middle section.}
    \label{fig:memoryCST}
\end{figure}

The device is electrically large so the full simulation is computationally expensive.
The experimental illuminating field is a Gaussian beam from a parabolic mirror with a diameter comparable to the cross-section of the device.
However, the simulation in \ac{CST} is done only for the central part of the device, as seen in Fig.~\ref{fig:memoryCST}, and the plane wave is considered as excitation. Due to the large width of the beam as compared to the target, the plane wave is an excellent approximation and simplifies the calculation.
Compared to the original geometry~\cite{Olejnik201AntiferromagneticMemory} the motive is made three times smaller. This shifts the frequency behavior to higher bands and results in improved performance of the excitation used.

The metric of interest is electric field intensity in CuMnAs layer~$E_\T{CuMnAs}$ in the center of the motive shown in Fig.~\ref{fig:memoryCST}.
The CuMnAs motive is prepared on a GaAs substrate and most of the CuMnAs alloy is covered by Au, except the central part as described in~\cite{Olejnik201AntiferromagneticMemory}. Incident field $E_\T{inc}$ is the excitation, which is represented by plane wave propagation orthogonally toward the motive, and polarization is aligned with two of the electrodes.
The delay from the source to the point of observation in CuMnAs is estimated to be $0.2 \, \T{ps}$.

\begin{figure*}
    \centering
    \includegraphics[]{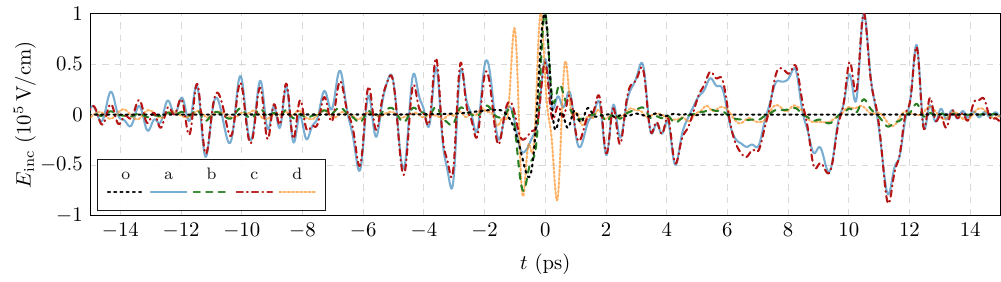}
    \caption{Time course of incident plane waves $E_\T{inc}$ at the position of the observation:
    (o) Pulse used in~\cite{Olejnik201AntiferromagneticMemory}.
    (a) Optimal excitation for the maximum peak.
    (b) Optimal excitation for the maximum peak with a constraint on the incident wave to concentrate $90\,\%$ of energy in the $2~\T{ps}$ interval.
    (c) Optimal excitation for the maximum peak with a constraint in the target field to concentrate $90\,\%$ of energy in the $14~\T{ps}$ interval.
    (d) Optimal excitation for the maximum peak with both constraints in (b) and (c).}
    \label{fig:inMemory}
\end{figure*}

\begin{figure*}
    \centering
    \includegraphics[]{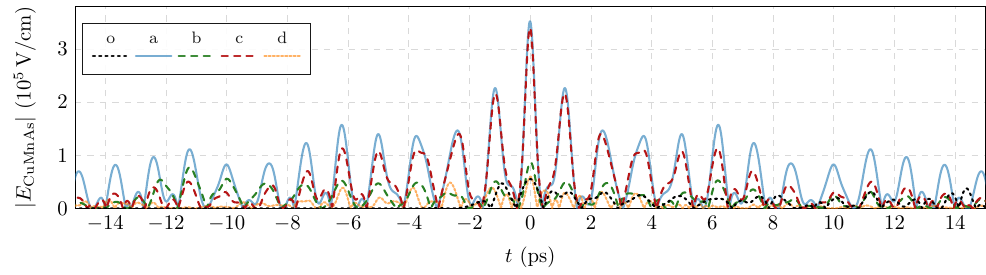}
    \caption{Magnitude of the electric field in CuMnAs  $E_\T{CuMnAs}$ for excitation pulses (o)--(d) in Fig.~\ref{fig:inMemory}.}
    \label{fig:outMemory}
\end{figure*}

The excitation waveforms are plotted in Fig.~\ref{fig:inMemory}.
The pulses are normalized to the maximum value of $10^5 \, \T{V/cm}$.
The excited field magnitude in CuMnAs $E_\T{CuMnAs}$ is provided in Fig.~\ref{fig:outMemory}.

The pulse used in~\cite{Olejnik201AntiferromagneticMemory} is denoted as (o).
Using the optimization framework within this paper, maximum switching performance is determined.
The objective function is the magnitude of the electric field intensity in CuMnAs.
The excitation is limited in frequency from $625\,\T{GHz}$ to $2.8\,\T{THz}$.
The constraint on the frequency band pertains to the frequency range of the excitation pulse (o).
Incident field energy is limited and excitation pulses are normalized to have a maximum amplitude of $10^5 \, \T{V/cm}$ as in~\cite{Olejnik201AntiferromagneticMemory}.

One of the optimal pulses presented corresponds to the maximum magnitude of the electric field in CuMnAs in $t = 0$ which is denoted as (a).
The optimization problem is identical to~\eqref{eq:maxE}, where $\M{q}$ corresponds to the expansion of incident plane wave $a (\omega)$ into the basis functions of Appendix~\ref{app:BF}, as done in~\eqref{eq:expan}.
The relation is computed using \ac{CST} which allows us to compute the electric field at a given point as a function of frequency considering unity excitation.
Mapping the set of the given basis functions results in the desired matrix.
The matrix representing the total energy of the incident plane wave is derived in Appendix~\ref{app:operators}. 

The maximum amplitude of the electric field in the CuMnAs alloy significantly exceeds that of the pulse used (o).
However, the excitation pulse has a long duration and is quite complex compared to (o).
Moreover, the resulting electric field in CuMnAs has many peaks and a long duration.

This evidence indicates the necessity of introducing constraints on the input and output waveforms.
The incident wave is constrained to have $90 \, \%$ energy in the interval $[-1.2,0.8]\, \T{ps}$ taking into account the delay of $t_\T{d} = 0.2 \, \T{ps}$.
The constraint is implemented using matrix derived from operator given in Appendix~\ref{app:operators}.
The corresponding \ac{QCQP} reads
\begin{equation}
    \begin{split}
        \max \limits_{\M{q}} \ & \M{q}^\herm \M{E}^\herm \M{E} \M{q}, \\
        \T{s.t.} \ & \M{q}^\herm \M{W} \M{q} = W_0, \\
        & \M{q}^\herm \left(0.9\M{W} - \M{W}_\T{T} \right) \M{q} \leq 0.
    \end{split}
    \end{equation}

The optimization solution is denoted as (b). The incident field waveform is now squeezed in the given narrow interval and is similar to the original pulse (o). However, the electric field in CuMnAs reaches a higher peak value

The same step with a time limitation can be done with the electric field in CuMnAs as follows
\begin{equation}
    \begin{split}
        \max \limits_{\M{q}} \ & \M{q}^\herm \M{E}^\herm \M{E} \M{q}, \\
        \T{s.t.} \ & \M{q}^\herm \M{W} \M{q} = W_0, \\
        & \M{q}^\herm \left(0.9\M{U} - \M{U}_\T{T} \right) \M{q} \leq 0,
    \end{split}
    \end{equation}
    where $\M{U}$ and $\M{U}_\T{T}$ are matrices of the total squared electric field and the squared electric field in the given time interval $T = 7 \, \T{ps}$ similarly derived as delivered energy matrices in Appendix~\ref{app:operators}.

The results are presented in (c) which shows the excitation again spanning a long time, whereas the waveform in CuMnAs outside of the prescribed interval is attenuated compared to the amplitude global maxima which are not far from the optimum without the constraint.

The combination of both constraints for excitation in (b) and for the target field in (c) results in another optimization problem whose solution is presented in (d).
Since optimization has more constraints than the previous solutions (a), (b), and (c), performance is lower but still higher than that of the original pulse (o).
That shows that optimization has achieved maximum potential in the given scenario considering the constraints.
The excitation mostly spans the narrow interval and the electric field in CuMnAs is also mainly accumulated in the given interval.
The disparity between (o) and (d) is minimal, so the used design performs almost the best possible. This suggests that enhancement can only be achieved through topological changes even more by co-design of the structure and of the pulse.

The current pulse has already reached its design limits.
Thus, achieving better performance necessitates either a redesigned system or a less restricted pulse.
This showcases the effectiveness of the tool in configuring pulses for antiferromagnetic memory switching and judging topologies for their capabilities.

\section{Optimization of the System's Topology}\label{sec:codesign}
Since the excitation waveform optimization is fast and robust, it can be combined with the topology optimization.
In this context, the focus is on the application of antenna far-field radiation. However, it should be noted that a similar degree of complexity can be applied to alternative applications, such as near-field focus or transmitter configurations, as soon as the transfer matrix~\eqref{eq:Etarget} is determined.
The antenna is a multiport system, and this example, as a by-product, proves its applicability to systems with multiple inputs\footnote{Multiple output systems within the objective can be treated by multiobjective optimization~\cite{Liska_etal_FundamentalBoundsEvaluation}, whose solutions represent Pareto optimal points.}.

The initial example consists of an array of two parallel and identical dipoles with the same dimensions as in  Section~\ref{sec:verification}. The frequency band also remains the same. The goal is to maximize instantaneous radiation in the end-fire direction.

The simplicity of the design enables rapid assessment of the impulse response and facilitates the transfer of data between iterations without requiring re-evaluation in a full-wave solver, which is beneficial when integrating system and waveform optimization. Application to larger structures with multiple ports does not bring any complications except for computation time.

\begin{figure}
    \centering
    \includegraphics{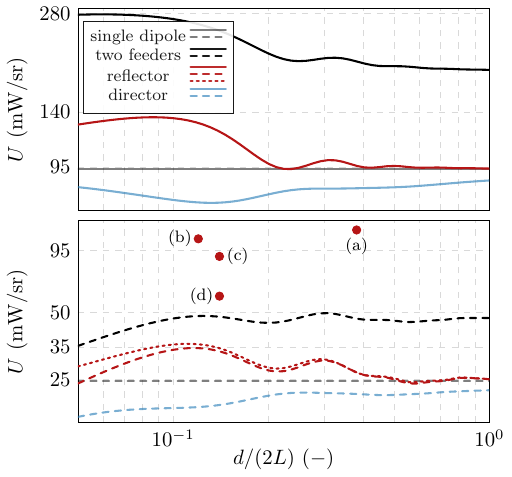}
    \caption{Comparison of maximum instantaneous radiation intensity $U$ for an end-fire two-dipole array for different distances between the dipoles normalized by the length of the dipoles. Performance is benchmarked against a single dipole. Considered configurations include both elements being driven, one element being driven with either a shorted director or reflector, and a reflector with an optimized load or a reflector setup with optimized topology. In the top panel, the solid lines show performance without considering impedance mismatch, reactive loading, or shape optimization, while, in the bottom panel, the dashed lines account for practical matching to an ideal $50\,\Omega$ feeding line, providing a more realistic assessment. The dotted line in the bottom panel represents a dipole reflector with an optimized reactive load, and the markers in the bottom panel represent reflector setups with optimized shapes~(a)-(d) in Fig.~\ref{fig:tsga}. Both optimizations in the bottom panel consider connection to ideal $50\,\Omega$ feed line.}
    \label{fig:distSweep}
\end{figure}

By performing a parametric sweep, the optimal spacing between the dipoles is determined. Different configurations are evaluated:
\begin{itemize}
    \item Both dipoles are driven.
    \item One dipole is driven, the other is shorted and acts as a reflector.
    \item One dipole is driven, the other is shorted and acts as a director.
    \item One dipole is driven and the reflector is, in its geometrical center, loaded by an optimized reactive loading.
\end{itemize}
The maximum instantaneous radiation intensity achieved by these setups is shown in Fig.~\ref{fig:distSweep} as a function of the distance between the dipoles, which is normalized to twice the dipole length, corresponding to the wavelength for which the dipole is designed.
\begin{figure}
    \centering
\begin{tabular}{|@{\,}c@{}@{}c@{\,}|}
\hline
\footnotesize{(a)} &  \\
    &  \includegraphics[height=0.8\linewidth, angle = 90]{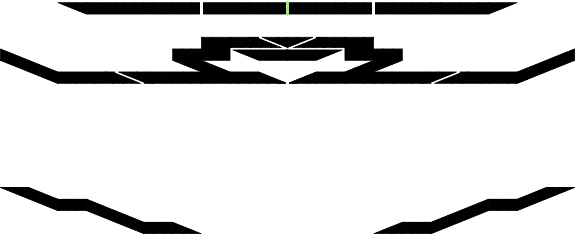} \\ \hdashline
    \footnotesize{(b)} &  \\
    &  \includegraphics[height=0.8\linewidth, angle = 90]{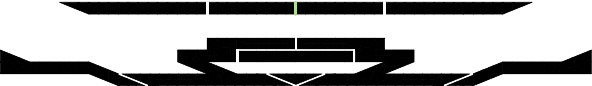} \\ \hdashline
    \footnotesize{(c)} &  \\
    &  \includegraphics[height=0.64\linewidth, angle = 90]{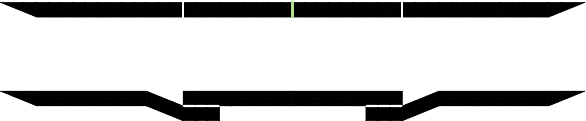}\\ \hdashline
    \footnotesize{(d)} &  \\
    &  \includegraphics[height=0.56\linewidth, angle = 90]{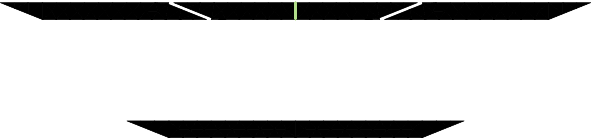} \\ \hline
\end{tabular}
\caption{Topologically optimized antenna shapes with ports highlighted in green. Instantaneous radiation intensity in the north direction is optimized (reflector setup):\\
    Shape (a) Best performance among the presented designs.\\
    Shape (b) Elements of the reflector closest to the driven element.\\
    Shape (c) Good trade-off between the shape regularity and performance.\\
    Shape (d) The simplest and the most compact design.}
    \label{fig:tsga}
\end{figure}

Figure~\ref{fig:distSweep} shows two panels:
\begin{itemize}
    \item In the top panel, the solid lines with better performance omit the matching (assume ideal voltage generators at all frequencies). The \acp{dof} are delta-gap voltages. In that case, the instantaneous far-field intensity is normalized to the total dissipated energy due to ohmic losses in the elements and radiation.
    \item  In the bottom panel, the dashed lines with lower performance consider the reflection of waves at the ports \ie, matching. The \acp{dof} are incident power waves in the $50\,\Omega$ feeding lines. There, the instantaneous far-field intensity is normalized to the total incident energy.
\end{itemize}

As expected, the best performance is achieved when both dipoles are fed independently. A similar trend is followed by a single fed dipole with a shorted reflector; for larger distances, the impact of the reflector is negligible, and the performance is identical to that of a dipole alone. The director decreases performance as compared to a single dipole.

Continuing in the co-design, the performance of the setup with reflector can be improved by loading the reflector with a shorted ideal $50 \,\Omega$ line of an optimized length $\ell$ in range from $0\,\T{m}$ to $2L$ (reactive loading). The optimization of reactive loading is only considered in the realistic scenario, where reflection caused by mismatch is taken into account, and the optimized \ac{dof} is the incident power wave. The enhancement is significant at small separations between the driven dipole and the reflector. The optimized design performance is plotted as a dotted line in Fig.~\ref{fig:distSweep} and can be compared to the dashed line for the shorted reflector. The best possible performance is $36.2\,\T{mW}$ which corresponds to $d/(2L) = 0.11$ and line length $\ell/(2L) = 0.20$. In the same way, the load on the director was also optimized, but the performance did not exceed that of a single dipole, so the result is not depicted.

The ultimate solution is co-design with the topology optimization~\cite{2021_capeketal_TSGAmemetics_Part1,2021_capeketal_TSGAmemetics_Part2}, which is implemented for the reflector setup. Performances are shown as markers in Fig.~\ref{fig:distSweep} for designs~(a)~--~(d) from Fig.~\ref{fig:tsga}, respectively. The antennas possess east-west mirror symmetry with respect to the north target direction. Designs are ranked according to their performance, inversely related to their regularity~\cite{Capek_etal_RegularityConstraints_EuCAP2021}. Performance enhancement is significant and underscores the effectiveness of the co-design methodology. This approach effectively addresses mismatch issues, improves gain, and exceeds the capabilities of reactive loading in terms of functionality.

\section{Conclusion} \label{sec:conclusion}
A methodology to determine optimal waveforms is presented in this paper with optimized parameters being defined as linear or quadratic terms.
The evaluation requires a post-processing step based on convex optimization and the input data can be calculated in any \ac{EM} simulator.
The technique can serve and provide quantitative results in many scenarios, including pulsed radiation or antiferromagnetic memory switching with pulse lasers, as discussed in this paper, as well as the creation of hotspots using near-field energy focusing in lossy materials using antenna arrays for potential applications in bioelectromagnetics.
Due to the generality inherent in the framework, along with the illustrative examples provided, the applicability extends to both the transmitting and receiving scenarios, without any increase in computational complexity once the system transfer matrix is supplied.

The capability of the tool is to find the optimal excitation waveform for an arbitrarily chosen target metric. Moreover, the number of sources is not restricted.
In the case of an antenna, matching can be omitted. This leads to fundamental bounds based on current density as known in the frequency domain, but now the current, apart from space, can be controlled in time allowing us to evaluate fundamental bounds in the time domain for a given design region based on the current, although the values would in many cases be overly optimistic.

An important strength of the proposed method lies in its practical utility for evaluating and optimizing radiating systems. Given a specific system configuration, the method determines whether the desired pulsed performance can be achieved.
If the shape and constraints of the system are sufficient, the method identifies the optimal pulse shape to maximize performance. Conversely, if the system cannot meet the imposed requirements, the approach highlights these limitations, offering valuable insights into whether a redesign of the system geometry is necessary.
By decoupling waveform optimization from structural design, the proposed method ensures a globally optimal solution that can be computed efficiently in polynomial time, unlike the iterative and often local solutions associated with topology optimization.
This characteristic guarantees the general applicability of the method to existing systems without necessitating their redesign, while still ensuring the best achievable performance for a given scenario and set of constraints.

The methodology has broader potential. It is extended to co-design scenarios in which both antenna and waveform design are considered simultaneously, making it applicable to a wider range of advanced engineering problems.
In such case, the topology optimization changes the design according to the results computed by the proposed tool. The combination of the pulse and topology design is present on an end-fire two-dipole array, resulting in a significant improvement of the performance.

\appendices
\section{Matrix Operators from \ac{EFIE} and \ac{MoM}} \label{app:MoM}
A representative example of an \ac{EM} solver capable of providing required antenna characteristics is \ac{AToM} based on \ac{EFIE} and \ac{MoM}.
The matrix representation of multiple quantities is presented in \cite[Appendix~A]{Liska_etal_FundamentalBoundsEvaluation}.
The extraction of essential metrics to the manuscript is presented in this appendix.
Multiple ports are considered.

Using \ac{EFIE} and \ac{MoM}, the electric field in the far-field zone expressed in the frequency domain reads
    \begin{equation}\label{eq:EtargetMoM}
        \V{E}_\infty (\V{r},\omega) = \Fvec(\V{r},\omega) \Ivec(\omega) \dfrac{\expn^{- \J \omega r/c}}{r},
    \end{equation}
    where $c$ is the speed of light, $\Fvec$ is a far-field matrix, and $\Ivec$ is a column vector of current density expansion coefficients.

The essential quantities are far field as a function of the incident waves from the feeding transmission line and their total available energy.
The electric far field is computed from far-field matrix~$\Fvec$ and current vector~$\Ivec$, see~\eqref{eq:EtargetMoM}.
The current is excited by incident waves~$\M{a}$ as
\begin{equation}
    \Ivec \left( \omega  \right) = \Zmat^{-1} \left( \omega  \right) \M{P} \M{v} \left( \omega  \right),
\end{equation}
where $\Zmat$ is the impedance matrix, and $\M{P}$ is the port matrix which assigns port voltage vector values~$\M{v}$ to corresponding positions for the matrix multiplication.
Port voltages are given by incident waves as
\begin{equation}
    \M{v} \left( \omega \right) = \M{k}_\T{i}^{-1} \left( \omega \right) \M{a} \left( \omega \right),
\end{equation}
with incident waves~$\M{a}$ and $\M{k}_\T{i}$ interconnecting the lossless feeding transmission lines of characteristic  impedance $Z_\T{char} = 50 \, \Omega$ considered in this manuscript with antenna port voltages~\cite{Capeketal_OptimalityOfTARCAndRealizedGainForMultiPortAntennas}.

Other antenna characteristics can be computed in the same way.
Emerging quantities for this application are antenna gain or port reflection coefficients.
The study of the characteristics provides information on the suitability of individual antennas for application.

\section{Energy and Field Operators} \label{app:operators}
Fourier transform is used to transform target quantities from the time domain to the frequency domain.
The mapping applies to energy and field expressions.

The field operator for a given location or direction, and time reads
\begin{equation}
    \V{E}(\V{r},t) = \\
    \int \limits_{-\infty}^\infty \M{H} \left( \V{r},  \omega  \right) \M{a} \left( \omega \right) \expn^{\J \omega t} \D \omega.
\end{equation}
One of the common targets is intensity~$\left| \V{E} \left(\V{r}, t \right)  \right|^2$  at optimized time $t_\T{opt}$. Choosing an arbitrary point in the space~$\V{r}$ and time window of length $2T$, another important measure is energy delivered to that point in that time window which reads
    \begin{multline} \label{eq:Wpart}
    W(\V{r}, T) = C \int \limits_{t_\T{d} - T }^{t_\T{d} + T } \left| \V{E} \left(\V{r}, t \right)  \right|^2 \T{d} t= \\
    C \int\limits_{\Omega } \int\limits_{\Omega } \dfrac{2 \sin\left[ \left(\omega_1 - \omega_2 \right) T \right] }{\omega_1 - \omega_2} \expn^{-\J \left( \omega_1 - \omega_2 \right) t_\T{d}(\V{r})} \\
    \M{a}^\herm \left( \omega_2  \right) \M{H}^\herm (\V{r},\omega_2) \M{H} (\V{r},\omega_1) \M{a} \left( \omega_1  \right) \T{d} \omega_2 \T{d} \omega_1,
\end{multline}
where $\Omega: \omega \in [-\omega_\T{max}, -\omega_\T{min}] \cup [\omega_\T{min}, \omega_\T{max}]$ is the intgration domain, $C$ is a constant to fit the units and $t_\T{d}$ is the time delay from the excitation source.

For incident energy, the transfer function~$\M{H}$ is omitted and the delay time is zero
\begin{multline}
    W_\T{in}(\V{r}, T) = C \int \limits_{t_\T{d} - T }^{t_\T{d} + T } \left| \M{a} \left( t \right)  \right|^2 \T{d} t= \\
 C \int\limits_{\Omega } \int\limits_{\Omega } \dfrac{2 \sin\left[ \left(\omega_1 - \omega_2 \right) T \right] }{\omega_1 - \omega_2} \M{a}^\herm \left( \omega_2  \right)  \M{a} \left( \omega_1  \right) \T{d} \omega_2 \T{d} \omega_1.
\end{multline}
The expression is simplified for the total energies, which read,
\begin{multline}
    W(\V{r}, T) = C \int \limits_{-\infty }^{\infty} \left| \V{E} \left(\V{r}, t \right)  \right|^2 \T{d} t= \\
    C \int\limits_{\Omega } \expn^{-\J \omega t_\T{d}(\V{r})} \M{a}^\herm \left( \omega \right) \M{H}^\herm (\V{r},\omega) \M{H} (\V{r},\omega) \M{a} \left( \omega  \right) \T{d} \omega,
\end{multline}
or
\begin{equation}
    W_\T{in}(\V{r}, T) = C \int \limits_{-\infty }^{\infty} \left| \M{a} \left( t \right)  \right|^2 \T{d} t= C \int\limits_{\Omega } \M{a}^\herm \left( \omega \right) \M{a} \left( \omega  \right) \T{d} \omega,
\end{equation}

Galerkin method~\cite{Kantorovich1982,Harrington_FieldComputationByMoM} converts continuous operators shown above to matrices using the same basis and testing functions.
The method is applied to all operators using functions introduced in Appendix~\ref{app:BF}.

\section{Band Limited Set of Basis Functions} \label{app:BF}
This manuscript uses a set of basis functions for the spectral expansion of antenna characteristics. 
Complex-valued expansion coefficients are denoted as~$q_n$.
The functions are band-limited, orthogonal, and continuous, and provide well-behaved real-valued images in~\ac{iFT}. Their explicit form reads
\begin{multline}
    \xi_n(\omega) = \dfrac{1}{\sqrt{\omega_\T{max} - \omega_\T{min}}} \\
    \sin{\left(n \pi \dfrac{|\omega| - \omega_\T{min}}{\omega_\T{max} - \omega_\T{min}} \right)}
    \begin{cases}
        1, \\
        \T{sign}\,(\omega)\J,
    \end{cases}
\end{multline}
where the upper row corresponds to the real part, while the lower row corresponds to the imaginary part. The values~$\omega_\T{min}$ and $\omega_\T{max}$ define the spectral band interval where the functions are nonzero.

\bibliographystyle{IEEEtran}
\bibliography{references.bib,TDbounds.bib}

\begin{thebibliography}{10}
\providecommand{\url}[1]{#1}
\csname url@samestyle\endcsname
\providecommand{\newblock}{\relax}
\providecommand{\bibinfo}[2]{#2}
\providecommand{\BIBentrySTDinterwordspacing}{\spaceskip=0pt\relax}
\providecommand{\BIBentryALTinterwordstretchfactor}{4}
\providecommand{\BIBentryALTinterwordspacing}{\spaceskip=\fontdimen2\font plus
\BIBentryALTinterwordstretchfactor\fontdimen3\font minus
  \fontdimen4\font\relax}
\providecommand{\BIBforeignlanguage}[2]{{%
\expandafter\ifx\csname l@#1\endcsname\relax
\typeout{** WARNING: IEEEtran.bst: No hyphenation pattern has been}%
\typeout{** loaded for the language `#1'. Using the pattern for}%
\typeout{** the default language instead.}%
\else
\language=\csname l@#1\endcsname
\fi
#2}}
\providecommand{\BIBdecl}{\relax}
\BIBdecl

\bibitem{1982_McIntosh_TAP}
R.~McIntosh and J.~Sarna, ``Bounds on the optimum performance of planar
  antennas for pulse radiation,'' \emph{{IEEE} Transactions on Antennas and
  Propagation}, vol.~30, no.~3, pp. 381--389, 1982.

\bibitem{1986_Kang_TAP}
Y.~Kang and D.~Pozar, ``Optimization of pulse radiation from dipole arrays for
  maximum energy in a specified time interval,'' \emph{{IEEE} Transactions on
  Antennas and Propagation}, vol.~34, no.~12, pp. 1383--1390, 1986.

\bibitem{Hackett2002}
D.~Hackett, C.~Taylor, D.~McLemore, H.~Dogliani, W.~Walton, and A.~Leyendecker,
  ``A transient array to increase the peak power delivered to a localized
  region in space: part {I}--theory and modeling,'' \emph{IEEE Transactions on
  Antennas and Propagation}, vol.~50, no.~12, pp. 1743--1750, Dec. 2002.

\bibitem{Maloney1993_OptimConicalAntPulseRad}
J.~Maloney and G.~Smith, ``Optimization of a conical antenna for pulse
  radiation: an efficient design using resistive loading,'' \emph{IEEE
  Transactions on Antennas and Propagation}, vol.~41, no.~7, pp. 940--947, Jul.
  1993.

\bibitem{Yang2005WallRadar}
Y.~Yang and A.~Fathy, ``See-through-wall imaging using ultra wideband
  short-pulse radar system,'' in \emph{2005 IEEE Antennas and Propagation
  Society International Symposium}, ser. APS-05.\hskip 1em plus 0.5em minus
  0.4em\relax IEEE, 2005.

\bibitem{Chen1999_LPDA}
Z.~Chen, M.~Foegelle, and T.~Harrington, ``Analysis of log periodic dipole
  array antennas for site validation and radiated emissions testing,'' in
  \emph{1999 IEEE International Symposium on Electromagnetic Compatability.
  Symposium Record (Cat. No.99CH36261)}, ser. ISEMC-99.\hskip 1em plus 0.5em
  minus 0.4em\relax IEEE, 1999.

\bibitem{1983_Pozar_TAP}
D.~Pozar, R.~McIntosh, and S.~Walker, ``The optimum feed voltage for a dipole
  antenna for pulse radiation,'' \emph{{IEEE} Transactions on Antennas and
  Propagation}, vol.~31, no.~4, pp. 563--569, 1983.

\bibitem{1984_Pozar_TAP}
D.~Pozar, D.~Schaubert, and R.~McIntosh, ``The optimum transient radiation from
  an arbitrary antenna,'' \emph{{IEEE} Transactions on Antennas and
  Propagation}, vol.~32, no.~6, pp. 633--640, 1984.

\bibitem{1985_Pozar_TAP}
D.~Pozar, Y.~Kang, D.~Schaubert, and R.~McIntosh, ``Optimization of the
  transient radiation from a dipole array,'' \emph{{IEEE} Transactions on
  Antennas and Propagation}, vol.~33, no.~1, pp. 69--75, 1985.

\bibitem{Onder1992TAP_OptControlFeedVoltageDipAntenna}
M.~Onder and M.~Kuzuoglu, ``Optimal control of the feed voltage of a dipole
  antenna,'' \emph{IEEE Transactions on Antennas and Propagation}, vol.~40,
  no.~4, pp. 414--421, Apr. 1992.

\bibitem{Olejnik201AntiferromagneticMemory}
K.~Olejník, T.~Seifert, Z.~Kašpar, V.~Novák, P.~Wadley, R.~P. Campion,
  M.~Baumgartner, P.~Gambardella, P.~Němec, J.~Wunderlich, J.~Sinova,
  P.~Kužel, M.~Müller, T.~Kampfrath, and T.~Jungwirth, ``Terahertz electrical
  writing speed in an antiferromagnetic memory,'' \emph{Science Advances},
  vol.~4, no.~3, Mar. 2018.

\bibitem{Etten1977}
\BIBentryALTinterwordspacing
P.~van Etten, ``The present technology of impulse radars,'' in \emph{Radar-77},
  Jan. 1977, pp. 535--539. [Online]. Available:
  \url{https://ui.adsabs.harvard.edu/abs/1977rpi..conf..535V}
\BIBentrySTDinterwordspacing

\bibitem{Polevoi_MaximumExtractableEnergy}
V.~G. Polevoi, ``Maximum energy extractable from an electromagnetic field,''
  \emph{Radiophysics and Quantum Electronics}, vol.~33, no.~7, pp. 603--609,
  1990.

\bibitem{Jelinek_UpperBoundOnInstantaneousPowerFluxAPS23}
L.~Jelinek, J.~Liska, and M.~Capek, ``Upper bound on instantaneous power
  flux,'' in \emph{2023 {IEEE} International Symposium on Antennas and
  Propagation and {USNC}-{URSI} Radio Science Meeting ({USNC}-{URSI})}.\hskip
  1em plus 0.5em minus 0.4em\relax {IEEE}, Jul. 2023.

\bibitem{Pozar2003_WaveformOptUWBRadioSys}
D.~Pozar, ``Waveform optimizations for ultrawideband radio systems,''
  \emph{IEEE Transactions on Antennas and Propagation}, vol.~51, no.~9, pp.
  2335--2345, Sep. 2003.

\bibitem{Liska2023}
J.~Liska, L.~Jelinek, and M.~Capek, ``Maximum peak radiation intensity,'' in
  \emph{2023 24th International Conference on Applied Electromagnetics and
  Communications (ICECOM)}.\hskip 1em plus 0.5em minus 0.4em\relax IEEE, Sep.
  2023.

\bibitem{GustafssonCapekSchab_TradeOffBetweenAntennaEfficiencyAndQfactor}
M.~Gustafsson, M.~Capek, and K.~Schab, ``Tradeoff between antenna efficiency
  and {Q}-factor,'' \emph{IEEE Trans. Antennas Propag.}, vol.~67, no.~4, pp.
  2482--2493, Apr. 2019.

\bibitem{GustafssonCapek_MaximumGainEffAreaAndDirectivity}
M.~Gustafsson and M.~Capek, ``Maximum gain, effective area, and directivity,''
  \emph{IEEE Trans. Antennas Propag.}, vol.~67, no.~8, pp. 5282 -- 5293, Aug.
  2019.

\bibitem{BoydVandenberghe_ConvexOptimization}
S.~Boyd and L.~Vandenberghe, \emph{Convex Optimization}.\hskip 1em plus 0.5em
  minus 0.4em\relax Cambridge, Great Britain: Cambridge University Press, 2004.

\bibitem{2012_Stutzman_Antenna_Theory}
W.~L. Stutzman and G.~A. Thiele, \emph{Antenna theory and design}.\hskip 1em
  plus 0.5em minus 0.4em\relax John Wiley \& Sons, 2012.

\bibitem{Pozar2007}
D.~M. Pozar, ``Optimal radiated waveforms from an arbitrary {UWB} antenna,''
  \emph{IEEE Transactions on Antennas and Propagation}, vol.~55, no.~12, pp.
  3384--3390, Dec. 2007.

\bibitem{Shlager1994_OptBowtieAntPulseRad}
K.~Shlager, G.~Smith, and J.~Maloney, ``Optimization of bow-tie antennas for
  pulse radiation,'' \emph{IEEE Transactions on Antennas and Propagation},
  vol.~42, no.~7, pp. 975--982, Jul. 1994.

\bibitem{Wang1997_OptimDipoleShapes}
J.-H. Wang, L.~Jen, and S.-S. Jian, ``Optimization of the dipole shapes for
  maximum peak values of the radiating pulse,'' in \emph{IEEE Antennas and
  Propagation Society International Symposium 1997. Digest}, ser. APS-97.\hskip
  1em plus 0.5em minus 0.4em\relax IEEE, 1997.

\bibitem{Pozar_MicrowaveEngineering}
D.~M. Pozar, \emph{Microwave Enginnering}, 4th~ed.\hskip 1em plus 0.5em minus
  0.4em\relax Wiley, 2011.

\bibitem{Wang2023}
X.-K. Wang, L.~Tian, H.-L. Wang, and X.-W. Zhu, ``Design of an ultra-wideband
  picosecond pulse generator based on step recovery diodes with an improved
  {SPICE} model,'' \emph{International Journal of Circuit Theory and
  Applications}, vol.~51, no.~8, pp. 3585--3595, Mar. 2023.

\bibitem{CST}
\BIBentryALTinterwordspacing
(2022) {CST Computer Simulation Technology}. Dassault Systemes. [Online].
  Available:
  \url{https://www.3ds.com/products-services/simulia/products/cst-studio-suite/}
\BIBentrySTDinterwordspacing

\bibitem{Kantorovich1982}
L.~V. Kantorovich and G.~P. Akilov, \emph{Functional analysis}.\hskip 1em plus
  0.5em minus 0.4em\relax Oxford New York: Pergamon Press, 1982.

\bibitem{Harrington_FieldComputationByMoM}
R.~F. Harrington, \emph{Field Computation by Moment Methods}.\hskip 1em plus
  0.5em minus 0.4em\relax Piscataway, New Jersey, United States: Wiley -- IEEE
  Press, 1993.

\bibitem{Liska-CompFunBoAntennas-EuCAP22}
J.~Liska, L.~Jelinek, and M.~Capek, ``Computation of fundamental bounds for
  antennas,'' in \emph{2022 16th European Conference on Antennas and
  Propagation ({EuCAP})}.\hskip 1em plus 0.5em minus 0.4em\relax {IEEE}, Mar.
  2022.

\bibitem{Liska_etal_FundamentalBoundsEvaluation}
\BIBentryALTinterwordspacing
------, ``Fundamental bounds to time-harmonic quadratic metrics in
  electromagnetism: {O}verview and implementation,'' \emph{arXiv}, 2021.
  [Online]. Available: \url{https://arxiv.org/abs/2110.05312}
\BIBentrySTDinterwordspacing

\bibitem{atom}
\BIBentryALTinterwordspacing
(2024) {A}ntenna {T}oolbox for {MATLAB} ({AToM}). Czech Technical University in
  Prague. {www.antennatoolbox.com}. [Online]. Available:
  \url{{www.antennatoolbox.com}}
\BIBentrySTDinterwordspacing

\bibitem{1999_Gustavsen_TPD}
B.~Gustavsen and A.~Semlyen, ``Rational approximation of frequency domain
  responses by vector fitting,'' \emph{IEEE Trans. Power Delivery}, vol.~14,
  pp. 1052--1061, 1999.

\bibitem{Yang2012GroundedBowtie}
J.~Yang and A.~Kishk, ``A novel low-profile compact directional ultra-wideband
  antenna: The self-grounded bow-tie antenna,'' \emph{IEEE Transactions on
  Antennas and Propagation}, vol.~60, no.~3, pp. 1214--1220, Mar. 2012.

\bibitem{Hertel2002ConicalSpiral}
T.~Hertel and G.~Smith, ``The conical spiral antenna over the ground,''
  \emph{IEEE Transactions on Antennas and Propagation}, vol.~50, no.~12, pp.
  1668--1675, Dec. 2002.

\bibitem{Balanis1989}
C.~A. Balanis, \emph{Advanced Engineering Electromagnetics}.\hskip 1em plus
  0.5em minus 0.4em\relax Hoboken, NJ: Wiley, 1989.

\bibitem{Balanis_Wiley_2005}
------, \emph{Antenna Theory Analysis and Design}, 3rd~ed.\hskip 1em plus 0.5em
  minus 0.4em\relax Wiley, 2005.

\bibitem{2021_capeketal_TSGAmemetics_Part1}
M.~Capek, M.~Gustafsson, L.~Jelinek, and P.~Kadlec, ``Optimal inverse design
  based on memetic algorithms -- {P}art {I}: {T}heory and implementation,''
  \emph{IEEE Trans. Antennas Propag.}, vol.~71, no.~11, pp. 8806--8816, 2021.

\bibitem{2021_capeketal_TSGAmemetics_Part2}
------, ``Optimal inverse design based on memetic algorithm -- {P}art {II}:
  {E}xamples and properties,'' \emph{IEEE Trans. Antennas Propag.}, vol.~71,
  no.~11, pp. 8817--8829, 2021.

\bibitem{Capek_etal_RegularityConstraints_EuCAP2021}
M.~Capek, V.~Neuman, J.~Tucek, L.~Jelinek, and M.~Gustafsson, ``Topology
  optimization of electrically small antennas with shape regularity
  constraints,'' in \emph{Proceedings of the 15th European Conference on
  Antennas and Propagation (EUCAP)}, 2021.

\bibitem{Capeketal_OptimalityOfTARCAndRealizedGainForMultiPortAntennas}
M.~Capek, L.~Jelinek, and M.~Masek, ``Finding optimal total active reflection
  coefficient and realized gain for multi-port lossy antennas,'' \emph{IEEE
  Trans. Antennas Propag.}, vol.~69, no.~5, pp. 2481--2493, Oct. 2021.

\end{thebibliography}

\begin{IEEEbiography}[{\includegraphics[width=1in,height=1.25in,clip,keepaspectratio]{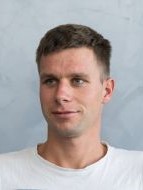}}]{Jakub Liska}
received his B.Sc. and M.Sc. degrees in electrical engineering from the Czech Technical University in Prague, Czech Republic, in 2019 and 2021, respectively. He is currently pursuing a Ph.D. in electrical engineering at the same university.

His research interests include \ac{EM} field theory, fundamental bounds, computational electromagnetics, numerical and convex optimization, numerical techniques, and eigenproblems.
\end{IEEEbiography}

\begin{IEEEbiography}[{\includegraphics[width=1in,height=1.25in,clip,keepaspectratio]{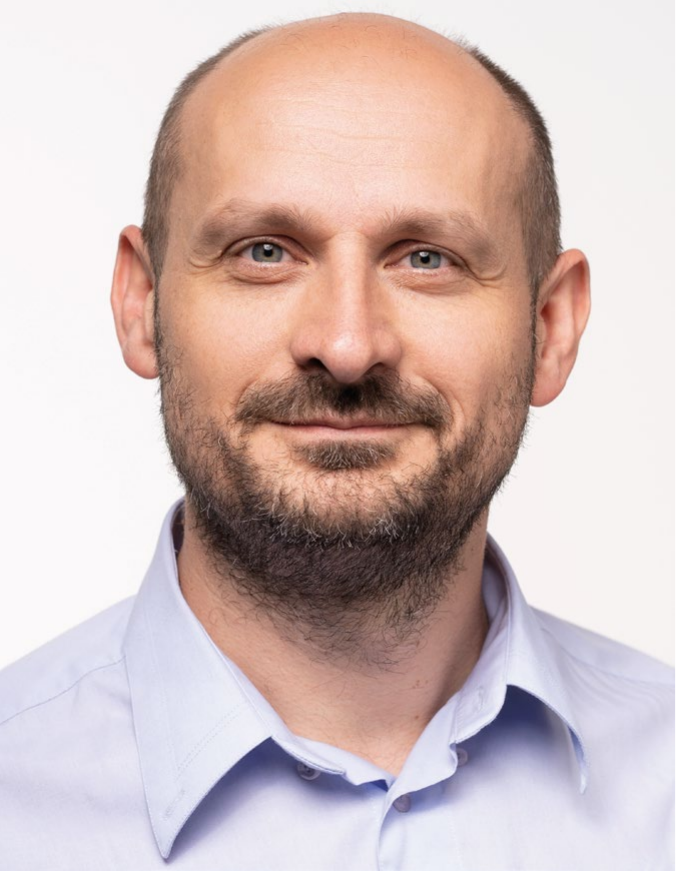}}]{Lukas Jelinek} was born in the Czech Republic in 1980. He received his Ph.D. from the Czech Technical University in Prague, Czech Republic, in 2006 for his work in metamaterials. In 2015, he received a permanent position at the Department of Electromagnetic Field at the same university.

His research interests include wave propagation in complex media, electromagnetic field theory, metamaterials, numerical techniques, and optimization.
\end{IEEEbiography}

\begin{IEEEbiography}[{\includegraphics[width=1in,height=1.25in,clip,keepaspectratio]{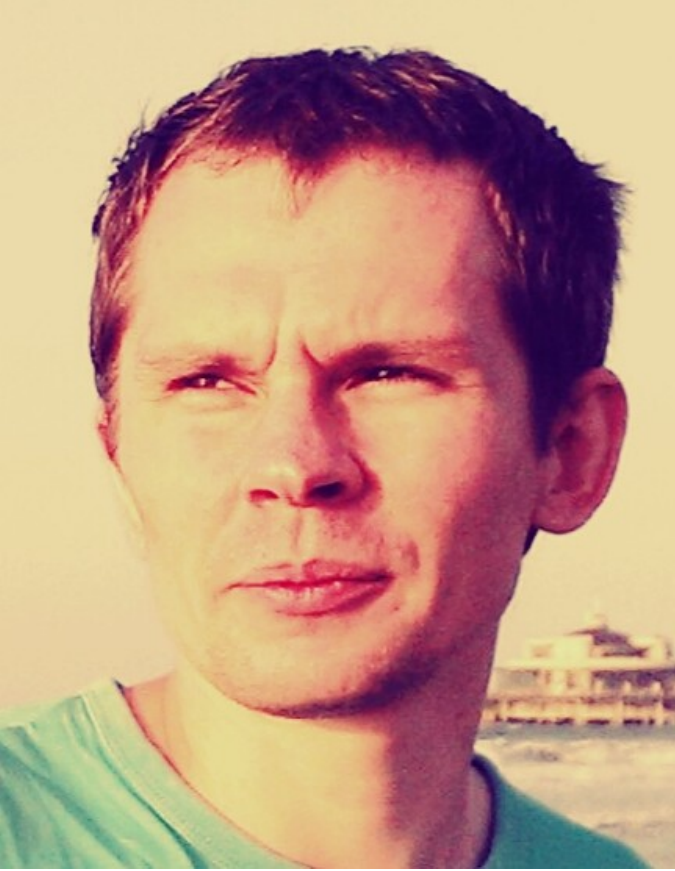}}]{Miloslav Capek}
(M'14, SM'17) received the M.Sc. degree in Electrical Engineering 2009, the Ph.D. degree in 2014, and was appointed a Full Professor in 2023, all from the Czech Technical University in Prague, Czech Republic.
	
He leads the development of the AToM (Antenna Toolbox for Matlab) package. His research interests include electromagnetic theory, electrically small antennas, antenna design, numerical techniques, and optimization. He authored or co-authored over 165~journal and conference papers.

Dr. Capek is the Associate Editor of IET Microwaves, Antennas \& Propagation. He was a regional delegate of EurAAP between 2015 and 2020 and an associate editor of Radioengineering between 2015 and 2018. He received the IEEE Antennas and Propagation Edward E. Altshuler Prize Paper Award~2022 and ESoA (European School of Antennas) Best Teacher Award in~2023.
\end{IEEEbiography}

\end{document}